# *N*-Way Joint Mutual Exclusion Does Not Imply Any Pairwise Mutual Exclusion for Propositions

**Roy S. Freedman**[1]

**Abstract**

Given a set of *N* propositions, if any pair is mutual exclusive, then the set of all propositions are *N*-way jointly mutually exclusive. This paper provides a new general counterexample to the converse. We prove that for any set of *N* propositional variables, there exist *N* propositions such that their *N*-way conjunction is zero, yet all *k*-way component conjunctions are non-zero. The consequence is that *N*-way joint mutual exclusion does not imply any pairwise mutual exclusion. A similar result is true for sets since propositional calculus and set theory are models for two-element Boolean algebra.

## 1   Introduction

Two propositions $X_1, X_2$ are mutually exclusive if they cannot be both true at the same time: the conjunction $X_1 \text{ and } X_2$ is never always true. The simplest example of this is the classical law of noncontradiction: the conjunction of $X$ and its negation $X'$ cannot both occur simultaneously. Even though one might at separate times observe $X$ to be true or false and $X'$ to be true or false, when both are observed together the result is always false. From a Boolean algebra perspective (denoting true by 1 and false by 0), propositions $X_1, X_2$ are mutually exclusive if the conjunction $X_1 X_2 = 0$. A proposition can be considered as a binary function of a set of *N* binary variables. This implies $X_1 = X_1(A_1, \cdots, A_N)$ and $X_2 = X_2(A_1, \cdots, A_N)$ are mutually exclusive if the conjunction $X_1 X_2 = 0$ for all possible combinations of binary variables $A_1, \cdots, A_N$.

In the case of three propositions $X_1, X_2, X_3$, mutual exclusion involves 3 pairs of conjunctions: $X_1 X_2$, $X_1 X_3$, and $X_2 X_3$. If any of the 3 pairs are mutually exclusive, then the joint triple conjunction $X_1 X_2 X_3$ is zero: $X_1, X_2, X_3$ cannot occur simultaneously. This is easily shown: using the associative and commutative rules of conjunction, we have

$$X_1 X_2 X_3 = (X_1 X_2) X_3 = X_1 (X_2 X_3) = (X_1 X_3) X_2 = 0$$

---

[1] Roy S. Freedman is with Inductive Solutions, Inc. and with the Department of Finance and Risk Engineering, New York University Tandon School of Engineering. Email: roy.freedman@nyu.edu.

In general, for *N* propositions, any pairwise mutual exclusion implies joint mutual exclusion. But is the converse true? Does joint *N*-way joint mutual exclusion imply pairwise mutual exclusion? Note that all we can say if $X_1 X_2 X_3 = 0$ is that proposition $X_1$ is pairwise exclusive of the conjunctive proposition $(X_2 X_3)$; proposition $X_2$ is pairwise exclusive of the conjunctive proposition $(X_1 X_3)$; and proposition $X_3$ is pairwise exclusive of the conjunctive proposition $(X_1 X_2)$. There is no requirement that any component conjunctive proposition $(X_j X_k)$ be zero.

This paper provides a new general counterexample: *N*-way joint mutual exclusion does not imply any pairwise mutual exclusion (or any *k*-way mutual exclusion for $2 \leq k < N$). The counterexample is constructed in the proof of the following result (which surprisingly has not appeared in the literature):

**Theorem A.** *Given a set of N propositional variables $A_1, A_2, ..., A_N$, there exist N non-zero propositions*

$$P_k = P_k(A_1, A_2, ..., A_N) \text{ with } P_k \neq 0, k = 1, 2, ..., N,$$

*having the following properties:*

*For $n = 2, ..., N-1$, the $\binom{N}{n}$ n-way component conjunctions:*

$$P_{k_1} P_{k_2} \cdots P_{k_n} \neq 0$$

*for $k_j = 1, 2, ..., N$, $k_1 < k_2 < ... < k_n = 1, 2, ..., N$, but the N-way joint conjunction*

$$P_1 P_2 \cdots P_N = 0.$$

In other words, for any *N* propositional variables, there exist *N* non-trivial propositions $P_k$ such that $P_1 P_2 \cdots P_N = 0$ yet any component conjunction of the $P_k$ are non-zero. The proof of Theorem A for $N \geq 4$, that involves the construction of the counterexample propositions $P_k$, is in Section 3.

Theorem A is true for sets of events (from the same sample space) since propositional calculus and set theory are models for two-element Boolean algebra. Recall that propositional variables in propositional calculus correspond to binary variables in two-element Boolean algebra and subsets in set theory; propositions correspond to binary functions. In propositional calculus, conjunction, disjunction, and negation correspond to set-theoretic intersection, union, and complement; and binary multiplication, and addition, and negation in two-element Boolean algebra. See [1, chapters 3,4,6] for a basic discussion of Boolean algebra, truth-functional (propositional) calculus and set theory; see Feller [2, Chapter 1] for the connections to probability.)

## 2  Notation and Examples

Out notation generally follows Hill and Peterson [1]. Recall that any $N$-ary proposition is a binary function $P_k = P_k(A_1, A_2, ..., A_N)$ with

$$P_k : \{0,1\}^N \rightarrow \{0,1\}.$$

Propositions can be represented as a truth table. The first $N$ columns of the table correspond to the domain of all $N$ binary variables (inputs); the rows corresponding to all combinations of the binary variables (this implies that there are $2^N$ rows). Column $N+1$ specifies the (output) value of the function for each binary variable combination. A minterm is the conjunction that corresponds to a single truth table input row. A binary function with several 1s in column $N+1$ corresponds to the disjunction of several minterms. Denote the disjunction $X_1 \, or \, X_1$ by $X_1 + X_2$. Every propositional truth function can be represented from its truth table as a disjunction of minterms: such a representation is the disjunctive normal form (or "sum of products") [5]. Note that all minterms are pairwise mutually exclusive.

We illustrate Theorem A for small $N$.

**Example: $N = 2, N = 3, N = 4$.** For two propositional variables $A_1, A_2$, consider the two propositions

$$P_1 = P_1(A_1, A_2) = A_1'A_2$$
$$P_2 = P_2(A_1, A_2) = A_1 A_2'.$$

Clearly $P_1 \neq P_2 \neq 0$ for all combinations of $A_1, A_2$. But the conjunction $P_1 P_2 = (A_1 A_1')(A_2 A_2') = 0$ for all combinations of $A_1, A_2$. For $N = 3$, three propositions that satisfy Theorem A are

$$P_1 = P_1(A_1, A_2, A_3) = A_2'A_3 + A_2 A_3'$$
$$P_2 = P_2(A_1, A_2, A_3) = A_1'A_3 + A_1 A_3'$$
$$P_3 = P_3(A_1, A_2, A_3) = A_1'A_2 + A_1 A_2'$$

All pairs of $P_k$ are not pairwise mutually exclusive, $P_i P_j \neq 0$; yet the joint 3-way conjunction is 3-way jointly mutual exclusive: $P_1 P_2 P_3 = 0$ (Figure 1).

|   | $A_1$ | $A_2$ | $A_3$ | $P_1$ | $P_2$ | $P_3$ | $P_1P_2$ | $P_1P_3$ | $P_2P_3$ | $P_1P_2P_3$ |
|---|---|---|---|---|---|---|---|---|---|---|
| 1 | 0 | 0 | 0 | 0 | 0 | 0 | 0 | 0 | 0 | 0 |
| 2 | 0 | 0 | 1 | 1 | 1 | 0 | 1 | 0 | 0 | 0 |
| 3 | 0 | 1 | 0 | 1 | 0 | 1 | 0 | 1 | 0 | 0 |
| 4 | 0 | 1 | 1 | 0 | 1 | 1 | 0 | 0 | 1 | 0 |
| 5 | 1 | 0 | 0 | 0 | 1 | 1 | 0 | 0 | 1 | 0 |
| 6 | 1 | 0 | 1 | 1 | 0 | 1 | 0 | 1 | 0 | 0 |
| 7 | 1 | 1 | 0 | 1 | 1 | 0 | 1 | 0 | 0 | 0 |
| 8 | 1 | 1 | 1 | 0 | 0 | 0 | 0 | 0 | 0 | 0 |

**Figure 1. Truth Table showing $P_j$ and all conjunctions for $N=3$.**

**Example: $N=4$.** The four propositions are

$$P_1 = A_2'A_3A_4 + A_2A_3'A_4 + A_2A_3A_4'$$
$$P_2 = A_1'A_3A_4 + A_1A_3'A_4 + A_1A_3A_4'$$
$$P_3 = A_1'A_2A_4 + A_1A_2'A_4 + A_1A_2A_4'$$
$$P_4 = A_1'A_2A_3 + A_1A_2'A_3 + A_1A_2A_3'$$

The six 2-way conjunctions are not pairwise mutually exclusive: $P_iP_j \neq 0$. The four 3-way conjunctions are not 3-way jointly mutually exclusive: $P_iP_jP_k \neq 0$. However, the single 4-way conjunction is jointly 4-way mutually exclusive: $P_1P_2P_3P_4 = 0$. These conjunctions for $N=4$ as shown in the Appendix.

## 3   Proof of Theorem A

Our proof is constructive. We have already shown the required $N$ propositions $P_k$ for $N = 2, 3, 4$. For $N \geq 4$, we show how to construct $N$ propositions $P_j$ such that the conjunction of any set of $N-1$ propositions is zero, but the conjunction of all $N$ propositions $P_j$ is zero:

$$\left. \begin{array}{l} P_{k_1} P_{k_2} \cdots P_{k_{N-1}} \neq 0, \ k_1 < k_2 < ... < k_{N-1} \\ P_1 P_2 \cdots P_N = 0 \end{array} \right\}. \tag{1}$$

First, construct the following $N$ "exclusion sets" $E_j$, defined by the set of all $N$ propositional variables $A_1, A_2, ..., A_N$ that exclude $A_j$ (denote exclusion of variable $A_j$ by $\hat{A}_j$). Each exclusion set has $N-1$ propositional variable elements. For $j = 1, 2, ..., N$,

$$E_j = \left\{ A_1, A_2, ...\hat{A}_j, ..., A_N \right\} = \left\{ A_1, A_2, ...A_{j-1}, A_{j+1}, ..., A_N \right\}. \tag{2}$$

Define an *e-term* for $E_j$ as a conjunction of all propositional variables in $E_j$ that has the form

$$A_1 A_2 \cdots \hat{A}_j \cdots A_k' \cdots A_N . \tag{3}$$

Each e-term is a conjunction of $N-1$ propositional variables from the exclusion set $E_j$ that excludes $A_j$ and contains just one single negation $A_k'$ of a propositional variable where $k \neq j$. For $E_j$, the set of all *e-terms* is

$$H_j = \left\{ A_1' A_2 \cdots \hat{A}_j \cdots A_N, A_1 A_2' \cdots \hat{A}_j \cdots A_N, \cdots, A_1 A_2 \cdots \hat{A}_j \cdots A_N' \right\} . \tag{4}$$

Construct propositions $P_j$ by forming the disjunction of all e-terms in $H_j$:

$$P_j = A_1' A_2 \cdots \hat{A}_j \cdots A_N + A_1 A_2' \cdots \hat{A}_j \cdots A_N + \ldots + A_1 A_2 \cdots \hat{A}_j \cdots A_N' \tag{5}$$

So

$$P_1 = A_2'A_3\ldots A_N + A_2 A_3'\ldots A_N + \ldots + A_2 A_3 \ldots A_j'\ldots A_N + A_2 A_3 \ldots A_N'$$
$$P_2 = A_1'A_3\ldots A_N + A_1 A_3'\ldots A_N + \ldots + A_1 A_3 \ldots A_j'\ldots A_N + A_1 A_3 \ldots A_N'$$
$$P_3 = A_1'A_2 A_4\ldots A_N + A_1 A_2' A_4 \ldots A_N + \ldots + A_1 A_2 A_3 \ldots A_j'\ldots A_N + A_1 A_2 A_4 \ldots A_N'$$
$$\ldots$$
$$P_N = A_1'A_2 A_3 \ldots A_{N-1} + A_1 A_2' A_3 \ldots A_{N-1} + A_1 A_2 A_3 \ldots A_j'\ldots A_{N-1} + \ldots + A_1 A_2 \ldots A_{N-1}'$$

Each $P_j$ excludes variable $A_j$. Every e-term in $P_j$ contains just one single negation of a propositional variable.

Now look at the conjunction of $N-1$ propositions $P_1 P_2 \cdots \hat{P}_j \cdots P_N$ and the conjunction $P_1 P_2 \cdots P_N$. For $N \geq 4$, it is convenient to arrange the e-terms of all $P_j$ in the tableau shown in Figure 2.

|       | 1 | 2 | … | j | … | N |
|-------|---|---|---|---|---|---|
| $P_1$ |   | $A_2'A_3\ldots A_N$ | … | $A_2 A_3 \ldots A_j'\ldots A_N$ | … | $A_2 A_3 A_4 \ldots A_N'$ |
| $P_2$ | $A_1'A_3\ldots A_N$ |   | … | $A_1 A_3 \ldots A_j'\ldots A_N$ | … | $A_1 A_3 A_4 \ldots A_N'$ |
| $P_3$ | $A_1'A_2 A_4\ldots A_N$ | $A_1 A_2' A_4 \ldots A_N$ | … | $A_1 A_2 A_4 \ldots A_j'\ldots A_N$ | … | $A_1 A_2 A_4 \ldots A_N'$ |
| $P_4$ | $A_1'A_2 A_3 A_5 \ldots A_N$ | $A_1 A_2' A_3 A_5 \ldots A_N$ | … | $A_1 A_2 A_3 A_5 \ldots A_j'\ldots A_N$ | … | $A_1 A_2 A_3 A_5 \ldots A_N'$ |
| … | … | … | … | … | … | … |
| $P_j$ | $A_1'A_2 \ldots A_{j-1} A_{j+1}\ldots A_N$ | $A_1 A_2' \ldots A_{j-1} A_{j+1}\ldots A_N$ | … |   | … | $A_1 A_2 \ldots A_{j-1} A_{j+1}\ldots A_N'$ |
| … | … | … | … | … | … | … |
| $P_N$ | $A_1'A_2 A_3 \ldots A_{N-1}$ | $A_1 A_2' A_3 \ldots A_{N-1}$ | … | $A_1 A_2 A_3 \ldots A_j'\ldots A_{N-1}$ | … |   |

**Figure 2. Tableau of e-terms of propositions $P_j$.**

Each row shows the e-terms of the Propositions $P_j$; the column label is the index of the negated propositional variable. The diagonal elements are all blank: there is no negated variable for $P_j$ in column $j$. This tableau allows us to easily consider the e-term by e-term conjunctions of the propositions. Denote $t(i, j)$ to be the e-term in row $i$ and column $j$. Note that an e-term $t(i, j)$ is blank only if it is on the tableau diagonal $i = j$.

We make the following observations:

1. **Two or more e-terms in the same column**: the conjunction of any two non-blank e-terms in a given column is non-zero: the conjunction $t(m, j)\, t(n, j)$ for $m \neq j$ and $n \neq j$ is

$$\left(A_1 A_2 \cdots \hat{A}_m \cdots A_j{}' \cdots A_N\right)\left(A_1 A_2 \cdots \hat{A}_n \cdots A_j{}' \cdots A_N\right) = A_1 A_2 \cdots A_j{}' \cdots A_N . \tag{6}$$

Note that the conjunction of all non-blank e-terms in same column yields the same result: for column $j$ and $m \neq j$, $t(1, j) \cdots t(m, j) \cdots t(N, j)$ for $m, n \neq j$ is

$$\left(A_1 A_2 \cdots \hat{A}_1 \cdots A_j{}' \cdots A_N\right) \cdots \left(A_1 A_2 \cdots A_j{}' \cdots A_N\right) = A_1 A_2 \cdots A_j{}' \cdots A_N . \tag{7}$$

2. **Two diagonally asymmetric e-terms**: the conjunction of any two non-blank e-terms $t(m, j)\, t(n, k)$ with $m \neq j$ and $n \neq j$ is zero provided $m \neq k$ and $n \neq j$:

$$\left(A_1 A_2 \cdots \hat{A}_m \cdots A_j{}' \cdots A_N\right)\left(A_1 A_2 \cdots \hat{A}_n \cdots A_k{}' \cdots A_N\right) = 0 . \tag{8}$$

3. **Two diagonally symmetric e-terms** $t(m, j) \wedge t(j, m)$ with $m \neq j$ and $n \neq j$ is non-zero and consists of two double negatives:

$$\left(A_1 A_2 \cdots \hat{A}_m \cdots A_j{}' \cdots A_N\right)\left(A_1 A_2 \cdots \hat{A}_j \cdots A_m{}' \cdots A_N\right) = A_1 A_2 \cdots A_j{}' \cdots A_m{}' \cdots A_N . \tag{9}$$

Note that conjunction of this double negative term with any other term in the tableau $\left(t(m, j)\, t(j, m)\right) t(n, k) = 0$ for $m \neq j$ and $n \neq k$, and where we exclude the original terms $(n, k) \neq (m, j), (j, m)$:

$$\left(A_1 A_2 \cdots A_j{}' \cdots A_m{}' \cdots A_N\right)\left(A_1 A_2 \cdots \hat{A}_n \cdots A_k{}' \cdots A_N\right) = 0 . \tag{10}$$

For $N \geq 4$, We claim that the conjunction of $N-1$ propositions $P_1 P_2 \cdots \hat{P}_j \cdots P_N$ is the conjunction of all terms in column $j$

$$P_1 P_2 \cdots \hat{P}_j \cdots P_N = A_1 A_2 \cdots A_j' \cdots A_N . \tag{11}$$

By observation 3, $(t(m,j)\,t(j,m))\,t(n,k) = 0$: the conjunction of three e-terms in three different rows is zero. By observation 2, $t(m,j)\,t(n,k) = 0$: the conjunction of two different e-terms in two different (asymmetric) rows is zero. So, the conjunction of at most three e-terms in three different rows is zero. However, observation 1 shows that the conjunction of two e-terms in the same column is

$$t(m,j)\,t(n,j) = A_1 A_2 \cdots A_j' \cdots A_N . \tag{12}$$

For $N \geq 4$, all "cross terms" drop out in the conjunction $P_1 P_2 \cdots \hat{P}_j \cdots P_N$, Finally, for the full $N$-term conjunction $P_1 P_2 \cdots P_N$

$$\begin{aligned} P_1 P_2 \cdots P_N &= \left( P_1 P_2 \cdots \hat{P}_j \cdots P_N \right)\left( P_1 P_2 \cdots \hat{P}_k \cdots P_N \right) \\ &= \left( A_1 A_2 \cdots A_j' \cdots A_N \right)\left( A_1 A_2 \cdots A_k' \cdots A_N \right) \\ &= 0 \end{aligned} \tag{13}$$

By the properties of conjunction, $X_1 X_2 \neq 0$ implies $X_1 \neq 0$ and $X_2 \neq 0$. Since the $N-1$ term conjunction $P_1 P_2 \cdots \hat{P}_j \cdots P_N \neq 0$, by induction, the $n$-term component conjunctions are also non-zero: $P_{k_1} \cdot P_{k_2} \cdot \ldots \cdot P_{k_n} \neq 0$. □

## 4   Discussion

We first illustrate Theorem A in a probability context of sets of events. Recall that two events $S_1$ and $S_2$ are mutually exclusive if the event intersection is the empty set: $S_1 S_2 = \emptyset$. Since $\Pr[\emptyset] = 0$, the probability of two pairwise mutually exclusive events $S_1, S_2$ occurring at the same time is $\Pr[S_1 S_2] = 0$. The joint conjunction of $N$ events is jointly $N$-way mutually exclusive if $S_1, S_2, \cdots S_N = \emptyset$ so that $\Pr[S_1, S_2, \cdots S_N] = 0$.

**Example: A Coin-Tossing Experiment.** Specify a binary coin and denote Head by 0 and Tail by 1. Given 3 coins, perform a triplet toss by throwing all three coins at the same time. Group the three coins of the triplet toss into three possible pairs. Each pair consists of three mutually exclusive events: zero Head, one Head, or two Heads. Let $S_k$ denote the event that coin-toss pair $k$ shows exactly one Head. If we are able to distinguish the coin triplet by $A,B,C$, then the three events are (see Figure 3):

$S_1$: Coin pair $(A,B)$ shows one Head;

$S_2$: Coin pair $(A,C)$ shows one Head;

$S_3$: Coin pair $(B,C)$ shows one Head.

| A | B |     | A | C |     | B | C |     |
|---|---|-----|---|---|-----|---|---|-----|
| 0 | 0 |     | 0 | 0 |     | 0 | 0 |     |
| 0 | 1 | $S_1$ | 0 | 1 | $S_2$ | 0 | 1 | $S_3$ |
| 1 | 0 |     | 1 | 0 |     | 1 | 0 |     |
| 1 | 1 |     | 1 | 1 |     | 1 | 1 |     |

**Figure 3. Triplet toss events for three coins grouped as three pairs.**

Events $S_1$, $S_2$, $S_3$ correspond to propositions $P_3$, $P_2$, $P_1$ with the three coins $A,B,C$ corresponding to $N=3$ propositional variables $A_1, A_2, A_3$. Figure 1 shows that $S_1$ and $S_2$ can occur simultaneously ($P_3 P_2 \neq 0$): we can observe a coin-toss pair $(A,B)$ showing one Head and a coin-toss pair $(A,C)$ showing one head. Similarly, $S_1$ and $S_3$ can occur simultaneously ($P_3 P_1 \neq 0$); $S_2$ and $S_3$ can occur simultaneously ($P_2 P_1 \neq 0$). Pairs of $P_k$ are not pairwise mutually exclusive: $P_i P_j \neq 0$ or, equivalently considering the intersection of sets of events, $S_i S_j \neq \varnothing$. Nevertheless, Theorem A shows that $S_1$, $S_2$, $S_3$ cannot jointly occur simultaneously since $P_1 P_2 P_3 = 0$, or equivalently, the 3-way intersection $S_1 S_2 S_3 = \varnothing$. The probability of $S_1 S_2 S_3$ occurring is zero.

In general, consider $N$ coins tossed at the same time and grouped into $N$ groups, each group containing $N-1$ coins. Let $S_j$ denote the event that coin-toss group $j$ shows exactly $N-2$ Heads: this corresponds with proposition $P_{N-j+1}$ as shown in equation (5). Theorem A shows that the joint intersection of all $N$ events $S_j$ is the null event: the $N$-way intersection $S_1 S_2 \cdots S_N = \varnothing$: we can never simultaneously see $N-2$ Heads appearing in all the $N$ groups of $N-1$ coins.

The observation that $N$-way joint mutual exclusion does not imply any pairwise or $k$-way joint mutual exclusion for propositions is similar to the observation that $N$-way joint independence does not imply pairwise independence for events. For example, three pairwise independent events from the same sample space, $S_1, S_2, S_3$ are pairwise independent if

$$\Pr[S_1 S_2] = \Pr[S_1]\Pr[S_2]; \quad \Pr[S_1 S_3] = \Pr[S_1]\Pr[S_3]; \quad \Pr[S_2 S_3] = \Pr[S_2]\Pr[S_3].$$

These three events are 3-way jointly independent if $\Pr[S_1 S_2 S_3] = \Pr[S_1] \cdot \Pr[S_2] \cdot \Pr[S_3]$. It is known since 1909 that pairwise independence does not imply *N*-way joint independence (see the counterexamples constructed by Bohlman (see Krengel [3, p.8]) and Lancaster [4]) . Conversely, *N*-way joint independence does not imply pairwise independence [5]. Several counterexamples of *N*-way joint independence versus pairwise independence have been collected by Khurshid and Sahai [6].

For future work we pose the question: For any *N*, are there other counterexamples concerning *N*-way joint mutual exclusion versus pairwise mutual exclusion?

# Appendix

| $A_1$ | $A_2$ | $A_3$ | $A_4$ | $P_1$ | $P_2$ | $P_3$ | $P_4$ | $P_1P_2$ | $P_1P_3$ | $P_1P_4$ | $P_2P_3$ | $P_2P_4$ | $P_3P_4$ | $P_1P_2P_3$ | $P_1P_2P_4$ | $P_1P_3P_4$ | $P_2P_3P_4$ | $P_1P_2P_3P_4$ |
|---|---|---|---|---|---|---|---|---|---|---|---|---|---|---|---|---|---|---|
| 0 | 0 | 0 | 0 | 0 | 0 | 0 | 0 | 0 | 0 | 0 | 0 | 0 | 0 | 0 | 0 | 0 | 0 | 0 |
| 0 | 0 | 0 | 1 | 0 | 0 | 0 | 0 | 0 | 0 | 0 | 0 | 0 | 0 | 0 | 0 | 0 | 0 | 0 |
| 0 | 0 | 1 | 0 | 0 | 0 | 0 | 0 | 0 | 0 | 0 | 0 | 0 | 0 | 0 | 0 | 0 | 0 | 0 |
| 0 | 0 | 1 | 1 | 1 | 1 | 0 | 0 | 1 | 0 | 0 | 0 | 0 | 0 | 0 | 0 | 0 | 0 | 0 |
| 0 | 1 | 0 | 0 | 0 | 0 | 0 | 0 | 0 | 0 | 0 | 0 | 0 | 0 | 0 | 0 | 0 | 0 | 0 |
| 0 | 1 | 0 | 1 | 1 | 0 | 1 | 0 | 0 | 1 | 0 | 0 | 0 | 0 | 0 | 0 | 0 | 0 | 0 |
| 0 | 1 | 1 | 0 | 1 | 0 | 0 | 1 | 0 | 0 | 1 | 0 | 0 | 0 | 0 | 0 | 0 | 0 | 0 |
| 0 | 1 | 1 | 1 | 0 | 1 | 1 | 1 | 0 | 0 | 0 | 1 | 1 | 1 | 0 | 0 | 0 | 1 | 0 |
| 1 | 0 | 0 | 0 | 0 | 0 | 0 | 0 | 0 | 0 | 0 | 0 | 0 | 0 | 0 | 0 | 0 | 0 | 0 |
| 1 | 0 | 0 | 1 | 0 | 1 | 1 | 0 | 0 | 0 | 0 | 1 | 0 | 0 | 0 | 0 | 0 | 0 | 0 |
| 1 | 0 | 1 | 0 | 0 | 1 | 0 | 1 | 0 | 0 | 0 | 0 | 1 | 0 | 0 | 0 | 0 | 0 | 0 |
| 1 | 0 | 1 | 1 | 1 | 0 | 1 | 1 | 0 | 1 | 1 | 0 | 0 | 1 | 0 | 0 | 1 | 0 | 0 |
| 1 | 1 | 0 | 0 | 0 | 0 | 1 | 1 | 0 | 0 | 0 | 0 | 0 | 1 | 0 | 0 | 0 | 0 | 0 |
| 1 | 1 | 0 | 1 | 1 | 1 | 0 | 1 | 1 | 0 | 1 | 0 | 1 | 0 | 0 | 1 | 0 | 0 | 0 |
| 1 | 1 | 1 | 0 | 1 | 1 | 1 | 0 | 1 | 1 | 0 | 1 | 0 | 0 | 1 | 0 | 0 | 0 | 0 |
| 1 | 1 | 1 | 1 | 0 | 0 | 0 | 0 | 0 | 0 | 0 | 0 | 0 | 0 | 0 | 0 | 0 | 0 | 0 |

**Figure 4.** Truth Table showing $P_j$ and all conjunctions for $N=4$.